\documentclass[12pt]{article}
\usepackage{a4wide}

\usepackage[dvipdfmx,hiresbb]{graphicx} %   usepackage with driver option
\usepackage{mediabb}                    %   [For PDF graphics] Automatic PDF-bounding box

\usepackage{amsmath,amssymb}
\usepackage{color}
\usepackage{tikz-feynman}
\usepackage{mathtools}
\usepackage{mathrsfs}
\usepackage{enumitem}
\usepackage[bookmarks,bookmarksnumbered]{hyperref}
\usepackage{cellspace}
\usepackage{tikz}
\usetikzlibrary{fadings}
\usetikzlibrary{patterns}
\usetikzlibrary{shadows.blur}
\usetikzlibrary{shapes}
\newcommand{\aln}[1]{\begin{align}#1\end{align}}
\newcommand{\nn}{\nonumber\\}

\begin{document}
\title{\vbox{
\baselineskip 14pt
\hfill \hbox{\normalsize 
}}  \vskip 1cm
\bf \Large On Virial Expansion in Hard Sphere Model
\vskip 0.5cm
}
\author{
Kiyoharu Kawana,\thanks{E-mail: \tt kkiyoharu@kias.re.kr}
\bigskip\\
\normalsize
\it $^\dagger$ School of Physics, Korean Institute for Advanced Study, Seoul 02455, Korea
\smallskip
}
\date{\today}

\maketitle   
%\vspace*{1cm}  
%%%%%%%%%%%%%%%%%%%% ABSTRACT %%%%%%%%%%%%%%%%%%
\begin{abstract} 
Virial expansion is a traditional approach in statistical mechanics that  expresses thermodynamic quantities, such as pressure $p$, as power series of density or chemical potential. 
Its radius of convergence can serve as a potential indicator of phase transition. 
In this study, we investigate the virial expansion of the hard-sphere model, using the known dimensionless virial coefficients $\tilde{B}_k^{}~(k=1,2,\cdots)$ up to the $12$th order. 
We find that it is well fitted by $\tilde{B}_k^{}=1.28\times k^{1.90}$, corresponding to the analytic continuation of the virial expansion of the pressure as $\sim \mathrm{Li}_{-1.90}^{}(\eta)$, where $\eta$ is the packing fraction and $\mathrm{Li}_s^{}(x)$ is the polylogarithm function.   
This implies the absence of singular behavior in the physical parameter space $\eta\leq \eta_{\mathrm{max}}^{}\approx 0.74$ and no indication of phase transition in the virial expansion approach.  
In addition, we calculate the cluster-integral coefficients $\{b_l^{}\}_{l=1}^\infty$ and observe that their asymptotic behavior resembles  the results obtained in the large dimension limit~($D\rightarrow \infty$), suggesting that $D=3$ might be already regarded as large dimension. 
However, the existence of phase transition in the hard-sphere model has been confirmed by numerous simulations, which clearly indicates that a naive extrapolation of the virial series can lead to unphysical   results.

\end{abstract} 

\setcounter{page}{1} 

\newpage  

%\tableofcontents   

%\newpage  

%____________________________________________
\section{Introduction}\label{Sec:intro}

Statistical mechanics of classical fluid systems has a long and rich  history. 
One of the traditional approaches is the virial expansion~\cite{mayer1977statistical,feynman1998statistical,hansen1990theory,Masters_2008} which represents thermodynamic quantities as  power series of density or chemical potential around the ideal gas. 
However, the convergence of the virial series is often slow, and it can  diverge at a finite density. 
Such a divergence can be interpreted as a potential indication of a phase transition although its occurrence is not necessarily guaranteed. 

One of the simplest yet non-trivial models in classical fluid systems is the hard-sphere model. 
Despite the absence of attractive forces, various numerical simulations~\cite{osti_4322875,Alder1959StudiesIM,PhysRev.127.359,Isobe_2015,doi:10.1142/10432} have consistently demonstrated that this system exhibits a phase transition within physical density region $\eta<\eta_{\mathrm{max}}^{}\approx 0.74$, where $\eta$ is the packing fraction. 
This intriguing result naturally motivates us to investigate the asymptotic behaviour of the virial series and its radius of convergence. 
%and see if it is close to the observed freezing density $\eta\sim 0.49$ if it exists.   
%
Indeed, there have already been a number of  studies~\cite{Baram_1979,10.1063/1.468456,Clisby_2005,B915002A,doi:10.1021/jp901041b,Oncak_2012}, employing various approximations and  resummation techniques. 

In this paper, we present an updated analysis of the asymptotic behavior of the virial expansion in the hard-sphere model, using the known virial coefficients up to the $12$th order~\cite{PhysRevLett.110.200601}.  
We find that these virial coefficients are well fitted by Eq.~(\ref{fitting 1}) and the estimated radius of convergence is $\eta_c^{}=1$, suggesting no phase transition within the physical density region $\eta<\eta_{\mathrm{max}}^{}$.  
Furthermore, the estimated virial coefficients correspond to the analytic continuation of the pressure as Eq.~(\ref{analytic continuation D=3}), and $\eta=\eta_c^{}=1$ is found to be a branch point of the polylogarithm function.     
We also calculate the cluster-integral coefficients $b_l^{}$ and observe that their asymptotic behavior resembles the results obtained in the large dimension limit~($D\rightarrow \infty$), suggesting that $D=3$ might be already regarded as large $D$. 
However, we should emphasize that it is still possible for the virial coefficients to become negative for $k>12$, which then can lead to a finite radius of convergence below $\eta_{\mathrm{max}}^{}\approx 0.74$.     

\

This paper is organized as follows. 
In Section~\ref{sec.2}, we provide a brief review the statistical mechanics of classical fluid system. 
We emphase that the existence of thermodynamic limit is guaranteed for a short-range potential that satisfies a couple of reasonable conditions. 
In Section~\ref{sec.3}, we discuss the applicability of virial expansion and present one qualitative way of estimating the radius of convergence in the large $l$ limit of the cluster integral $\{b_l^{}(V)\}_{l=1}^{\infty}$. 
In Section~\ref{sec.4}, we investigate the virial expansion in the hard-sphere model. 
For comparison, we first review $D=1$ and $\infty$ limits, where the model is exactly solvable. 
We then examine the $D=3$ case, estimating the asymptotic behavior of the virial series using the known virial coefficients up to the $12$th order.  
Conclusion is given in Section~\ref{sec:conclusion}.

\newpage

%________________
\section{Classical fluid system and thermodynamic limit}\label{sec.2}
%\subsection{Partition function and thermodynamic limit}
We consider thermodynamic equilibrium of classical particles interacting with a Hamiltonian in the $D$-dimensional Euclidean space:
\aln{
H_N^{}=\sum_{i=1}^N\frac{p_i^2}{2m}+U_N^{}(\{x_i^{}\}_{i=1}^N)~,\quad U_N^{}(\{x_i^{}\}_{i=1}^N)=\sum_{i<j}^Nv(x_i^{},x_j^{})~,
\label{def:Hamiltonian}
}
where $x_i^{}=(x_i^1,x_i^2,\cdots,x_i^D)$, $p_i^{2}=\sum_{M=1}^D (p_i^M)^2$, and $v(x_i^{},x_j^{})$ is a general two-body potential.    
If we impose the translation invariance, the pair potential becomes a mere function of relative coordinate as $v(x_1^{},x_2^{})=v(x_1^{}-x_2^{})$, and we focus on this case in the following. 
%\footnote{ 
%
%Moreover, if we impose the rotation symmetry $O(d)$, the pair potential is further restricted to be $v(x)=v(|x|)$.
%
%This is the simplest class of models of classical gas and called simple gas/liquid.  
% 
%}

%
The grand-canonical partition function is defined by
\aln{
\Xi [T,\mu,V]
 &=\sum_{N=0}^\infty \frac{1}{N!}\left(\prod_{i=1}^N\int_V d^Dx_i^{}\int_V d^Dp_i^{}\right)e^{
-\beta H_N^{}+\beta \mu N}
\nn
&=\sum_{N=0}^\infty \frac{z^N}{N!}\left(\prod_{i=1}^N\int_V d^Dx_i^{}\right)e^{-\beta\sum_{i<j}v(x_{i}^{}-x_j^{})}~,
\label{grand-canonical partition function}
}
where $\beta=1/T$ is the inverse temperature, $\mu$ is the chemical potential, and $z=e^{\beta\mu}(2\pi mT)^{D/2}$ is the fugacity.   
Here, $\sum_{i<j}$ means the summation over all different pairs $(x_i^{},x_j^{})$ among $N$ particles.   
The integrand in Eq.~(\ref{grand-canonical partition function}) can be written as 
\aln{
W_N^{}(\{x_i^{}\}_{i=1}^N)&\coloneq e^{-\beta\sum_{i<j}v(x_{i}^{},x_j^{})}=\prod_{i<j}(1+f_{ij}^{})
\nn
&=1+\sum_{i<j}f_{ij}^{}+\sum_{i<j}\underset{(k,l)\neq (i,j)}{\sum_{k<l}}f_{ij}^{}f_{kl}^{}+\cdots~,
}
where $f_{ij}^{}\coloneq e^{-\beta v(x_i^{}-x_j^{})}-1$ is the Mayer's $f$ function. 
Then, the cluster expansion allows us to perform the summation of $N$ in Eq.~(\ref{grand-canonical partition function}) as~\cite{mayer1977statistical,feynman1998statistical}
%it can be shown that the summation of $N$ can be performed in Eq.~(\ref{grand-canonical partition function}) as~\cite{}
\aln{
\log \Xi[T,\mu,V]=V\sum_{l=1}^{\infty}b_l^{}(V)z^l~,
\label{cluster expansion}
}   
where 
\aln{
b_k^{}(V)\coloneq \frac{1}{k!V}\left(\prod_{i=1}^{k}d^Dx_i^{}\right)C_k^{}(\{x_i^{}\}_{i=1}^k)~
}
is the cluster integral. 
Here, $C_k^{}(\{x_i^{}\}_{i=1}^k)$ denotes the contribution in $W_k^{}(\{x_i^{}\}_{i=1}^k)$ such that all the $k$-particles are connected each other by the Mayer's function, i.e. connected diagrams.   
In addition, the density is similarly expressed as 
\aln{
\frac{\langle N\rangle}{V}=\frac{\partial}{\partial (\beta \mu)}\frac{\log \Xi[T,\mu,V]}{V}=\sum_{l=1}^{\infty}lb_l^{}(V)z^l~.
}

The existence of the thermodynamic limit 
%It is non-trivial The fundamental question is the grand-potential or pressure in the thermodynamic limit is defined by 
\aln{
\beta p(T,\mu)&\coloneq \lim_{V\rightarrow \infty}\frac{\log \Xi[T,\mu,V]}{V}~
\label{thermodynamic limit}
%\\
%&=\sum_{k=1}^{\infty}b_k^{}(\infty)z^k~.
}
is a nontrivial problem for a generic $N$-body potential; 
however, in classical fluid systems, it is rigorously proven for short-range $N$-body potentials that satisfy several reasonable conditions~\cite{ruelle1999statistical}.  
Since this is a crucial result, we briefly overview it  below. 
%
%Note that the following discussion applies not only for the pair potential~(\ref{def:Hamiltonian}), but also for a general $N$-body potential $U_N^{}(\{x_i^{}\}_{i=1}^N)=\sum_{i<j}v(x_i^{}-x_j^{})+\sum_{i<j<k}v(x_i^{},x_j^{},x_k^{})+\cdots$. 
 
We first define the mutual potential energy by
\aln{
M_{N_1^{}N_2^{}}(\{x_i^{}\}_{i=1}^{N_1^{}};\{x_j^{'}\}_{j=1}^{N_2^{}})\coloneq U_{N_1^{}+N_2^{}}(\{x_i^{}\}_{i=1}^{N_1^{}},\{x_j^{'}\}_{j=1}^{N_2^{}})-U_{N_1^{}}(\{x_i^{}\}_{i=1}^{N_1^{}})-U_{N_2^{}}^{}(\{x_j^{'}\}_{j=1}^{N_2^{}})~. 
}
Then, we say that $U_N^{}(\{x_i^{}\}_{i=1}^N)$ is {\it tempered} if there exist $^\exists \lambda >D$ and $^\exists R_0^{}>0$, $^\exists A\geq 0$ such that 
\aln{
M_{N_1^{}N_2^{}}(\{x_i^{}\}_{i=1}^{N_1^{}};\{x_j^{'}\}_{j=1}^{N_2^{}})\leq AN_1^{}N_2^{}e^{-\lambda r}
}
whenever $|x_i^{}-x_j^{'}|\geq r\geq R_0^{}$ for all $i=1,2,\cdots, N_1^{}$ and $j=1,2,\cdots, N_2^{}$. 
Roughly speaking, this temperedness implies the suppression of the  positive part of $U_N^{}$ for the long-distance $r\rightarrow \infty$.  
For example, in the pair potential case~(\ref{def:Hamiltonian}), this implies (with $N_1^{}=N_2^{}=1$)
\aln{
v(x)\leq A |x|^{-\lambda}\quad \text{for}~|x|\geq R_0^{}~,
}
which implies that positive (repulsive) part of $v(x)$ should damp faster than $|x|^{-\lambda}(<|x|^{-D})$ for $|x|\rightarrow \infty$.   

In addition, we say that $U_N^{}(\{x_i^{}\}_{i=1}^N)$ is {\it stable} if there exists $^\exists B\geq 0$ such that  
\aln{
U_N^{}(\{x_i^{}\}_{i=1}^N)\geq -NB
}
for all $N=1,2,\cdots$ and $^\forall (x_1^{},\cdots,x_N^{})\in (\mathbb{R}^D)^N$. 
The importance of this stability is that it implies the convergence of the grand-canonical partition function as
\aln{
\Xi[T,\mu,V]&=1+\sum_{N=1}^\infty \frac{z^N}{N!}\left(\prod_{i=1}^N\int_V d^Dx_i^{}\right)e^{-\beta U_N^{}(\{x_i^{}\}_{i=1}^N)}
\nn
&\leq 1+\sum_{N=1}^\infty \frac{z^N}{N!}\left(\prod_{i=1}^N\int_V d^Dx_i^{}\right)e^{-N\beta B}
\nn
&=\exp\left(Vze^{\beta B}\right)~
}
for $^\forall z\geq 0$ and $^\forall \beta\geq 0$. 
Here, we list typical pair potentials that satisfy temperedness and stability~\cite{ruelle1999statistical}:
\\
\\
$\bullet$ {\bf Hard sphere potential}
\aln{
v_{\mathrm{HS}}^{}(x)=\begin{cases} \infty & \text{for} |x|<2a
\\
0 & \text{for} |x|\geq 2a
\end{cases}~,
\label{hard sphere potential}
}
where $a>0$ corresponds to the radius of molecule.
\\
\\
$\bullet$ {\bf Pair potential with hard core}\\
More generally, we can consider
\aln{
v(x)=v_{\mathrm{HS}}^{}(x)+\delta v(x)~,
}
where the positive part of $\delta v(x)$ decreases faster than $|x|^{-D}$ for $|x|\rightarrow \infty$ and satisfies 
\aln{\sum_{i=2}^N \delta v(x_j^{}-x_1^{})\geq -B
} 
for all $N$ and $^\forall (x_1^{},x_2^{},\cdots ,x_N^{})$ such that $|x_i^{}-x_1^{}|\geq 2a$ for $^\forall i\neq 1$.  
\\
\\ 
$\bullet$ {\bf Lennard-Jones potential}
\aln{
v(x)=v_0^{}\left[\left(\frac{\sigma}{r}\right)^p-\left(\frac{\sigma}{r}\right)^q\right] 
}
with $p\geq D$ and $q\geq D+1$.   

\

\noindent Now the existence theorem of the thermodynamic limit~(\ref{thermodynamic limit}) is given as follows.

\

\noindent {\it Theorem}. The thermodynamic limit~(\ref{thermodynamic limit}) exists for a stable and tempered potential $U_N^{}(\{x_i^{}\}_{i=1}^N)$. 
$\beta p(\mu,T)$ is a convex function of $\mu$ and $T$.  
In particular, $p(\mu,T)$ is a continuous function of $(\mu,T)$ and an increasing function of $\mu$.   

\

\noindent In the next section, we will argue that this rigorous result clearly highlights the applicability of cluster/virial expansion exclusively to a single unique phase.

%______________________________________________________
\section{Cluster/Virial expansion and its applicability}\label{sec.3}

%______________________________________________________
\begin{figure}[t!]
\begin{center}
\includegraphics[scale=0.7]{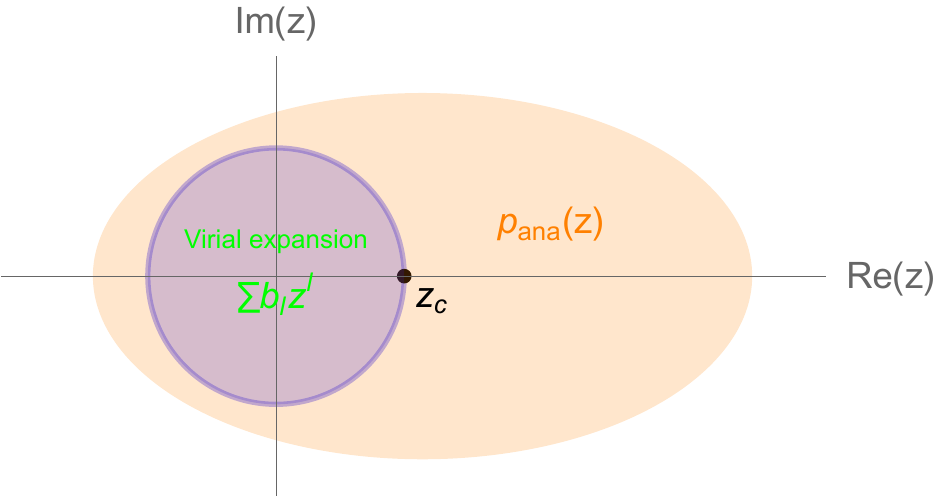}
\caption{
Virial expansion and its analytic continuation.
We may find an analytic continuation $p_{\mathrm{ana}}^{}(z)$ of virial series, but it cannot represent the true pressure $p(z)$ beyond $z=z_c^{}$ when a system exhibits a phase transition at this point.    
}
\label{fig:zplane}
\end{center}
\end{figure}
%______________________________________________________

We have seen that the grand-canonical partition function is given by  Eq.~(\ref{cluster expansion}) based on cluster expansion. 
The existence of thermodynamic limit implies that the pressure and density can be written as 
\aln{
\beta p(T,\mu)=\sum_{l=1}^{\infty}b_l^{}(\infty)z^l~,\quad  n(T,\mu)=\frac{\partial (\beta p)}{\partial \log z}=\sum_{l=1}^{\infty}lb_l^{}(\infty)z^l~.
\label{asymptotic series of pressure}
}
However, the convergence of these series is not necessarily   guaranteed for $^\forall z\geq 0$. 
If the radius of convergence is finite and given by $|z|=z_c^{}<\infty$, it may be possible to construct an analytic continuation $p_{\mathrm{ana}}(z)$ of the virial series, whose domain contains some interval of the positive real axis $z\geq z_c^{}$, as illustrated in Fig.~\ref{fig:zplane}. 
Then, there are typically three possibilities:
%______________________________
\begin{enumerate}[label=(\alph*)]
\item %$\lim_{z\rightarrow z_c^{}} p_{\mathrm{ana}}^{}(z)=\infty$ or indefinite, i.e. 
$z=z_c^{}$ is a pole or essential singularity of $p_{\mathrm{ana}}^{}(z)$.   
\item %$\lim_{z\rightarrow z_c^{}} p_{\mathrm{ana}}^{}(z)$ 
$z=z_c^{}$ is a regular point of $p_{\mathrm{ana}^{}}(z)$.   
\item 
%$\lim_{z\rightarrow z_c^{}} p_{\mathrm{ana}}^{}(z)<\infty$, but 
$z=z_c^{}$ is a branch point of $p_{\mathrm{ana}}^{}(z)$.  
\end{enumerate}
%______________________________
In case (a), $p_{\mathrm{ana}}^{}(z)$ cannot represent the true pressure for $z\geq z_c^{}$, as the singular behavior at $z=z_c^{}$ is inconsistent with the theorem. 
Thus, it can only describe the equilibrium of classical fluid for $z<z_c^{}$, i.e. gas phase. 
On the other hand, the case (b) can be consistent with the theorem, and $p_{\mathrm{ana}}^{}(z)$ can represent the true pressure even for $z\geq z_c^{}$. 
In this case, there is no singularity at $z=z_c^{}$ and the system stays in an unique phase throughout the domain of $p_{\mathrm{ana}}^{}(z)$.  
The case (c) is subtle and may also be consistent with the theorem if $\lim_{z\rightarrow z_c^{}}p_{\mathrm{ana}}^{}(z)<\infty$ and $p_{\mathrm{ana}}^{}(z)$ is real and positive for $z\geq z_c^{}$ in a certain branch. 

In any cases, the cluster/virial expansion~(\ref{asymptotic series of pressure}) can only describe a single phase and cannot account for phase transition~\cite{10.1063/1.1750208,PhysRevLett.109.040601}. 
%

%the right-hand side in Eq.~(\ref{asymptotic series of pressure}) may diverge  for $z\rightarrow z_c^{}$, but this is clearly  inconsistent with the rigorous theorem in the previous section.  
%
%In this sense, the cluster expansion method is only applicable within the radius of convergence (i.e. gas    phase) and cannot explain phase transition~\cite{}. 
%
%This situation is depicted in Fig.~\ref{}.
%
%We should note that even if the convergence radius is $|z|=z_c^{}<\infty$, the series in Eq.~(\ref{asymptotic series of pressure}) can be finite at $z=z_c^{}$ when $b_l^{}(V)$ is an alternating series, $b_l^{}(\infty)=(-1)^{l+1}a_l^{}~,a_l^{}\geq 0$.
%
%In such a case, it might be possible to perform analytic continuation of the right-hand side for $z>z_c^{}$, and system indicates no singular behaviors at $z=z_c^{}$.  
%
%We will see that the cluster/virial expansion in hard-sphere model belongs to this case. 

\

Nevertheless, the estimation of radius of convergence offers a potential indication of phase transition or critical point.  
One qualitative way to estimate the convergence radius is the steepest decent method in the $l\rightarrow \infty$ limit, based on the irreducible cluster integrals.  
In general, $b_k^{}(V)$ can be expressed as 
\aln{
l^2b_l^{}(V)=\sum_{\{n_k^{}\}}\prod_{k=1}^{l-1}\frac{(l \beta_k^{}(V))^{n_k^{}}}{n_k^{}!}~,
\label{cluster by irreducible cluster}
}
where $\sum_{\{n_k^{}\}}$ is the summation over all combinations of $(l-1)$ positive integers $\{n_k^{}\}_{i=1}^{l-1}$ satisfying $\sum_{k=1}^{l-1}kn_k^{}=l-1$, and $\beta_k^{}(V)$ denotes the irreducible contribution in $b_k^{}(V)$. 
By the residue theorem, Eq.~(\ref{cluster by irreducible cluster}) is also  expressed as 
\aln{
l^2b_l^{}(V)=\frac{1}{2\pi i}\oint_C d\xi \frac{e^{l\sum_{k=1}^\infty \beta_k^{}(V)\xi^k}}{\xi^l}~,
\label{bl by contour integral}
} 
where $C$ is an anticlockwise contour around $\xi=0$ in the complex $\xi$ plane.  
We denote the integrand as 
\aln{
e^{l\sum_{k=1}^\infty \beta_k^{}(V)\xi^k}/\xi^l=e^{lF(\xi)}~,
}  
where
\aln{
F(\xi)=\sum_{k=1}^{\infty}\beta_k^{}(V)\xi^k-\log \xi~. 
%\simeq \sum_{k=1}^{\infty}\beta_k^{}\xi^k -\log \xi~.
}
One can check that a saddle point exists at the point $\xi_0^{}\in \mathbb{R}$ satisfying\footnote{
This statement is always true when $\beta_k^{}(V)\geq 0$ for $^\forall k$ while it is not when $\beta_k^{}(V)$ can take negative values. 
} 
\aln{
\frac{dF(\xi)}{d\xi}\bigg|_{\xi=\xi_0^{}}^{}=0\quad \leftrightarrow \quad \sum_{k=1}^{\infty} k\beta_k^{}(V)\xi_0^{k}=1~,
\label{saddle point}
}
and we can modify the contour $C$ so that it passes this saddle point. 
Then, $F(\xi)$ is expanded around the saddle point as 
\aln{F(\xi)=F(\xi_0^{})+\frac{F''(\xi_0^{})}{2}(\xi-\xi_0^{})^2+{\cal O}((\xi-\xi_0^{})^3)~,
}
where 
\aln{
F''(\xi_0^{})&=\sum_{k=2}^{\infty} \beta_k^{}(V)k(k-1)\xi_0^{k-2}+\frac{1}{\xi_0^{2}}=\frac{1}{\xi_0^2}\sum_{k=2}^{\infty}k^2 \beta_k^{}(V)\xi_0^k~.
}
which can be positive or negative depending on $\beta_k^{}(V)$.
When it is negative, this already indicates the instability of system at the density $n=\xi_0^{}$ because the series $\sum_{k=1}^{\infty}k^2\beta_k^{}(V)n^k$ corresponds to the second-oder derivative of the grand-canonical potential with respect to the density $n$. (See Eq.~(\ref{Virial expression}) below.)  
Thus, it suffices to consider the positive case, and the direction of steepest decent is given by 
\aln{
\xi-\xi_0^{}=ix~,\quad x\in \mathbb{R}~,
%\begin{cases} ix 
%& \text{for $F''(\xi_0^{})> 0$}
%\\
%x & \text{for $F''(\xi_0^{})<0$}
%\end{cases},\quad x\in \mathbb{R}~,
} 
and the saddle point approximation of Eq.~(\ref{bl by contour integral}) is 
\aln{
l^2b_l^{}(V)\simeq \frac{1}{\sqrt{2\pi lF''(\xi_0^{})}}e^{lF(\xi_0^{})}\quad \therefore~b_l^{}(V)\simeq \frac{c}{l^{5/2}}b_0^{l}~,
\label{bl by steepest descent}
}
where
\aln{b_0^{}=e^{F(\xi_0^{})}=\frac{1}{\xi_0^{}}e^{\sum_{k=1}^{\infty}\beta_k^{}(V)\xi_0^k}~,\quad c=\frac{1}{\sqrt{2\pi F''(\xi_0^{})}}~,
}
and $\xi_0^{}$ is determined by Eq.~(\ref{saddle point}). 
Now, we can estimate the radius of convergence $z_c^{}$ of the series~(\ref{cluster expansion}) as 
\aln{z_c^{}=\left(\lim_{l\rightarrow \infty} b_l^{}(V)^{1/l}\right)^{-1}=b_0^{-1}=\xi_0^{}e^{-\sum_{k=1}^{\infty}\beta_k^{}(V)\xi_0^k}~.
}
Namely, the radius of convergence is determined by the irreducible cluster integrals $\{\beta_k^{}(V)\}_{k=1}^\infty$ in the large $l$ limit. 
Note that these irreducible cluster integrals also allow us to express the  pressure as a power series of density as~\cite{mayer1977statistical,10.1063/1.1750208} 
\aln{
\sum_{l=1}^{\infty} b_l^{}(V)z^l=n\left(1-\sum_{k=1}^{\infty}\frac{k}{k+1}\beta_k^{}(V)n^{k}\right)~,\quad n=ze^{-\sum_{k=1}^{\infty}\beta_k^{}(V)z^k}~,
\label{Virial expression}
}
which is the popular form of the Virial expansion.
%and this implies that we obtain the same radius of convergence in both series. 
%  
%

%_________________________________
\section{Hard Sphere Model} \label{sec.4}

We study the asymptotic behavior of the virial expansion of the hard-sphere model~(\ref{hard sphere potential}). 
Note that, the cluster/virial coefficients do not depend on temperature $T$ because the Mayer's $f$ function behaves as  
\aln{f(x)=e^{-\beta v(x)}-1=\begin{cases} -1 & |x|<2a 
\\
0 & |x|>2a 
\end{cases}
}
in the hard-sphere model. 

\subsection{$D=1$ and $\infty$}
The hard-sphere model is exactly solvable for $D=1$ and $\infty$. 

\

\noindent \underline{$D=1$}\\
When $D=1$, the canonical partition function can be evaluated as (with $x_{N+1}^{}=L$)
\aln{
\frac{Z_N^{}(L)}{N!}&=\left(\prod_{i=1}^{N}\int_0^{x_{i+1}^{}}dx_i^{}\right)\prod_{i=1}^{N-1}e^{-\beta \sum_{i=1}^{N-1}v(x_{i+1}^{}-x_i^{})}
\nn
&=\frac{1}{N!}\left(\int_0^{L-2Na} dxe^{-\beta v(x)}\right)^{N}
\nn
&=\frac{1}{N!}(L-2Na-2a)^N~.
\label{canonical partition function}
%\int_{a}^{L-Na}dx_1^{}\int_{x_1^{}+a}^{L-(N-1)a}\cdots \int_{}^{} 
}
%As a slight generalization of Eq.~(\ref{hard sphere potential}), let us consider 
%\aln{
%v(x)=\begin{cases} +\infty & 0\leq x<2a
%\\
%-v_0^{} & 2a\leq x<2(a+b)
%\\
%0 & 2(a+b)\leq x \leq L
%\end{cases}
%}
%with $0<b<a$. 
%
%In this case, Eq.~(\ref{canonical partition function}) is evaluated as 
%\aln{
%\frac{Z_N^{}(L)}{N!}=\frac{1}{N!}\left(2be^{\beta v_0^{}}+(L-Na-2(a+b))\right)^{N}~.
%}
By using the Stirling's formula $N!\sim \sqrt{2\pi N}(N/e)^N$, we obtain
\aln{
Z_N^{}(L)\sim \sqrt{2\pi (N-1)}\left(\frac{L}{N}-2a
\right)^{N}~. 
}
Then, the pressure is given by
\aln{
\beta p a=\lim_{N\rightarrow \infty}a\frac{\partial}{\partial L}\log Z_N^{}(L)= \frac{\eta}{1-\eta}~,\quad \eta=\frac{2Na}{L}~.
}
In this case, the pressure diverges at $\eta=\eta_{\mathrm{max}}^{}=1$, and there is no phase transition in the physical density region $\eta<1$.  

\

\noindent \underline{$D=\infty$}\\
For $D\rightarrow \infty$, the evaluation of the cluster integral $b_l^{}(V)$ is extremely simplified because only the connected tree-diagrams survive~\cite{PhysRevA.36.2422}. 
Namely, it is 
\aln{\lim_{D\rightarrow \infty}b_l^{}(V)=\frac{l^{l-2}}{l!}(-v)^{l-1}~,
}
where $l^{l-2}$ is the number of connected tree diagrams with $l$ distinguished particles (Cayler's theorem), and  
\aln{
-v=\lim_{D\rightarrow \infty}\int d^Dx f(x)=\lim_{D\rightarrow \infty}V_D^{}(a)~,\quad V_D^{}(a)=\frac{\pi^{D/2}a^{D}}{\Gamma(\frac{D}{2}+1)}~.
}
Here, $V_D^{}(a)$ is the volume of $D$-dimensional ball.  
We assume that this limit is finite.  

Now, the pressure and density are given in cluster expansion as 
\aln{\beta p v&=v\sum_{l=1}^{\infty}b_l^{}(V)z^l=\sum_{l=1}^{\infty}(-1)^{l+1}\frac{l^{l-2}}{l!}(vz)^{l}~,
\label{large D pressure}
\\
nv&=v\sum_{l=1}^{\infty}lb_l^{}(V)z^l=\sum_{l=1}^{\infty}(-1)^{l+1}\frac{l^{l-1}}{l!}(vz)^{l}~,
\label{large D density}
}
and both series have the same radius of convergence
\aln{vz_c^{}=\left(\lim_{l\rightarrow \infty}\left(\frac{l^l}{l!}\right)^{\frac{1}{l}}
\right)^{-1}=e^{-1}~.
}
In this case, we can find approximate analytic continuations by using the Stirling's formula as
\aln{\beta p v\approx -\frac{1}{\sqrt{2\pi}}\sum_{l=1}^{\infty}\frac{1}{l^{\frac{5}{2}}}(-ezv)^l=-\frac{1}{\sqrt{2\pi}}\mathrm{Li}_{\frac{5}{2}}(-evz)~,
\label{p in large D}
\\
n v\approx -\frac{1}{\sqrt{2\pi}}\sum_{l=1}^{\infty}\frac{1}{l^{\frac{3}{2}}}(-ezv)^l=-\frac{1}{\sqrt{2\pi}}\mathrm{Li}_{\frac{3}{2}}(-evz)~,
\label{n in large D}
} 
where $\mathrm{Li}_s^{}(x)$ is the polylogarithm function defined by analytic continuation of 
\aln{
\mathrm{Li}_s^{}(x)=\sum_{k=1}^{\infty}\frac{x^n}{k^s}~,
}
which is holomorphic except for a branch cut $x\geq 1$ in the real axis. 
This implies that Eqs.~(\ref{p in large D})(\ref{n in large D}) provide approximate pressure and density of the system in the entire physical parameter space $z\geq 0$ in the large $D$ limit. 
Apparently, there is no phase transition.  

The equation of state is much simpler compared to the above expansions in terms of $z$. 
In the large $D$ limit, all the irreducible cluster integrals $\beta_k^{}(V)$ for $k\geq 2$ are negligible because they contain loops. 
Thus, only $\beta_1^{}(V)$ survives, and we have 
\aln{
\beta p v=\eta+\frac{1}{2}\eta^2~,\quad \eta=vn~,
}
which is nothing but the mean-field result.

\subsection{$D=3$}

Now let us consider $D=3$. 
In this case, only the first few virial coefficients are known.  
The virial expansion in the right-hand side in  Eq.~(\ref{Virial expression}) is often written as
\aln{
\beta p=\sum_{k=1}^{\infty}B_k^{}n^k~,\quad B_k^{}=-\frac{k-1}{k}\beta_{k-1}^{}~,\quad B_1^{}=1~.
}
In the hard-sphere model, it is conventional to express this expansion in terms of dimensionless density as  
\aln{
\beta pV_3^{}(a)=\sum_{k=1}^{\infty}\tilde{B}_k^{}(T)\eta^k~,\label{dimensionless virial}
} 
where 
\aln{
\tilde{B}_k^{}(T)\coloneq \frac{B_k^{}(T)}{V_3^{}(a)^{k-1}}~,\quad \eta \coloneq V_3^{}(a)n~,
}
Here, $\eta$ is known as the packing fraction and its maximum value in $D=3$ is $\eta_{\mathrm{max}^{}}\approx 0.74$. 
% 
%______________________________________________________
\begin{figure}[t!]
\begin{center}
\includegraphics[scale=0.5]{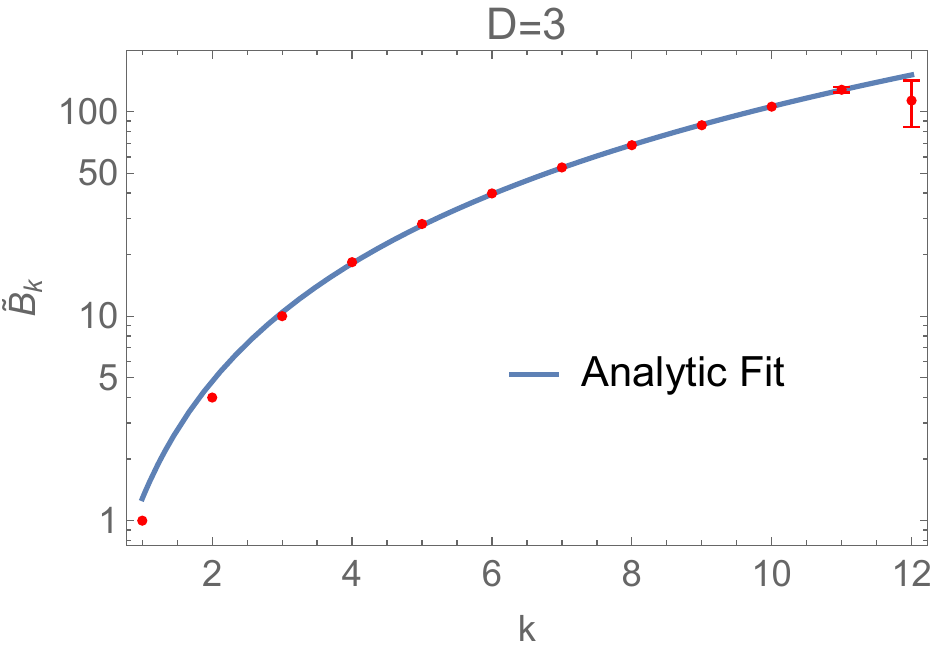}
\includegraphics[scale=0.5]{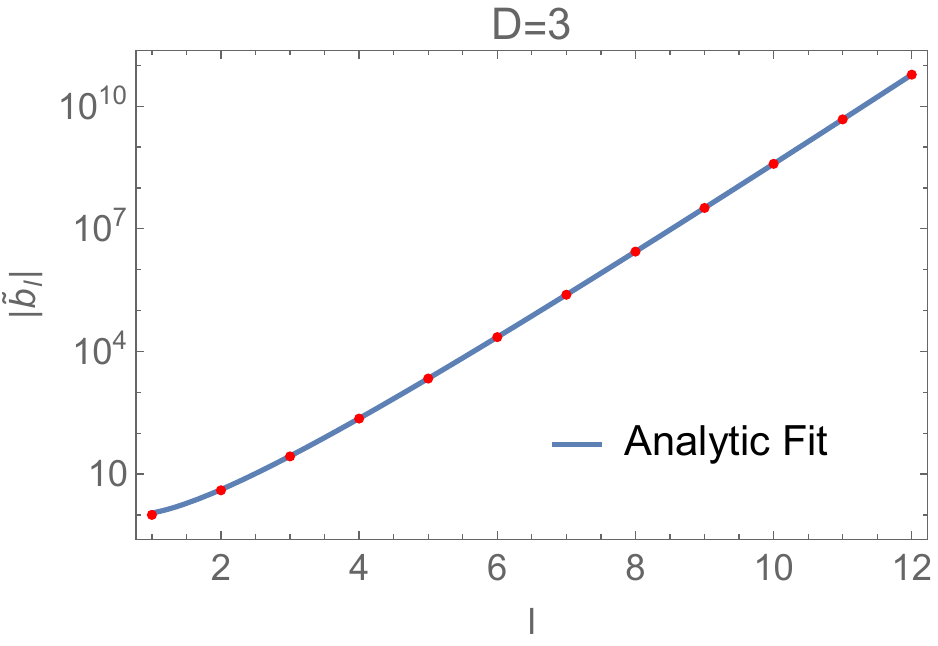}
\includegraphics[scale=0.5]{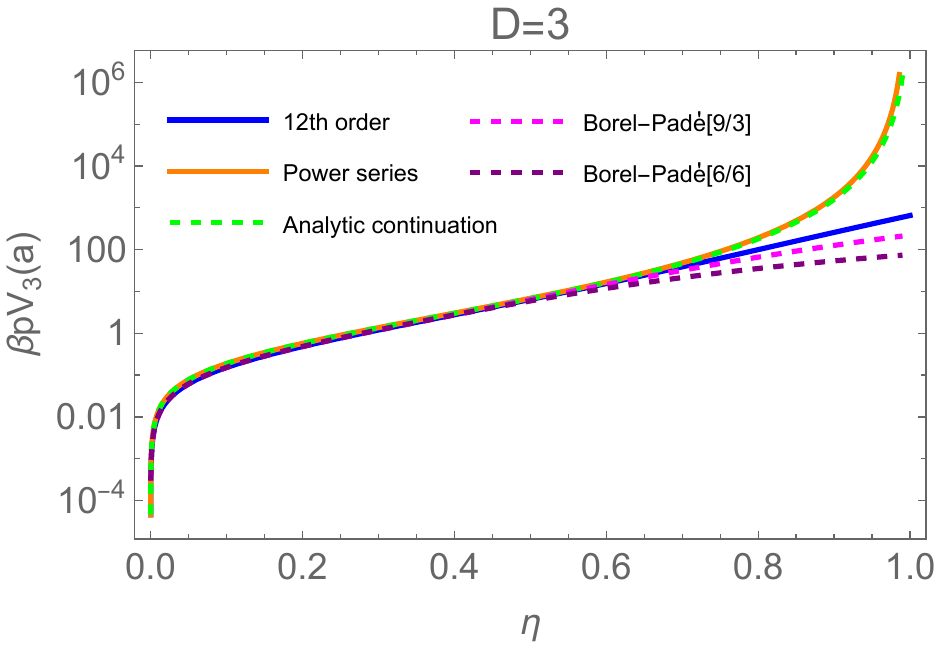}
\includegraphics[scale=0.5]{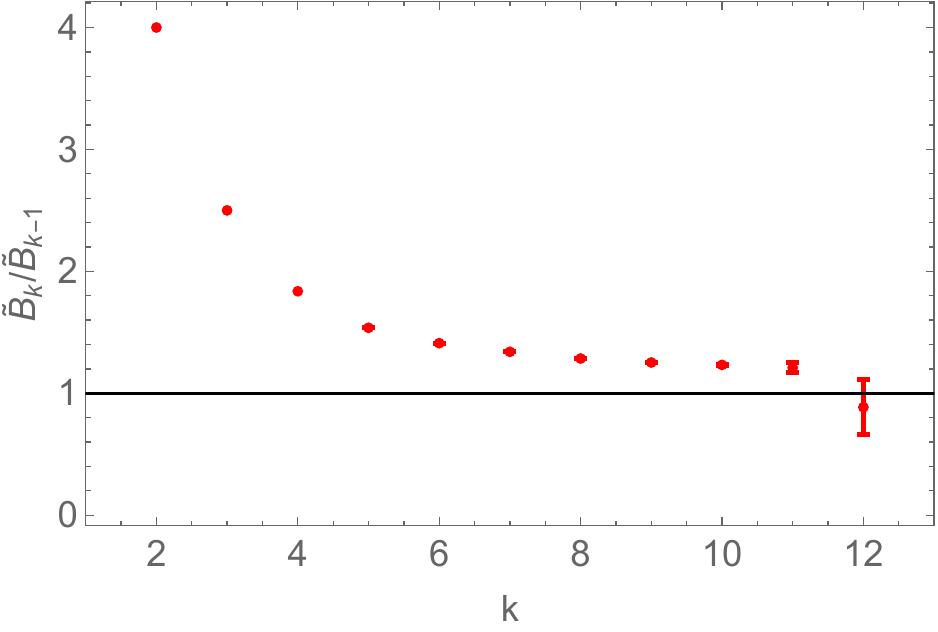}
\caption{Upper Left (Right): Virial coefficients (cluster-integral coefficients) of the hard sphere model in $D=3$. 
\\
Lower Left: Pressure as a function of the packing fraction: blue for the virial expansion up to $12$th order, orange for the predicted infinite power series, dashed green for its analytical continuation~(\ref{analytic continuation D=3}), dashed magenta (purple) for the Borel-Pad\'e resummation with $[9/3]~([6,6])$. 
\\
Lower Right: Plot of the ratio $\tilde{B}_k^{}/\tilde{B}_{k-1}^{}$. 
}
\label{fig:virial}
\end{center}
\end{figure}
%______________________________________________________

These virial coefficients are analytically calculated up to the fourth order~\cite{Boltzmann,PhysRev.85.777,Lyberg_2005}:
\aln{
\tilde{B}_2^{}=4~,\quad \tilde{B}_3^{}=10~,\quad \tilde{B}_4^{}=\frac{2707\pi+438\sqrt{2}-4131\arccos(1/3)}{70\pi}=18.3647684~.
} 
The higher-order virial coefficients are numerically calculated up to the $12$th order~\cite{Clisby_2005,PhysRevLett.110.200601}:
\aln{
&\tilde{B}_5^{}=28.2245(3)~,\quad \tilde{B}_6^{}=39.81514(93)~,\quad \tilde{B}_7^{}=53.3444(37)~,
\nn
&\tilde{B}_8^{}=68.537(18)~,\quad \tilde{B}_9^{}=85.812(85)~,\quad \tilde{B}_{10}^{}=105.77(39)~,
\label{known virial coefficients}
\\
&\tilde{B}_{11}^{}=127.9(42)~,\quad \tilde{B}_{12}^{}=113(29)~.
\nonumber
}
One can check that these coefficients are well fitted by the following  function:
\aln{
\tilde{B}_k^{\mathrm{fit}}=1.28\times k^{1.90}~. 
\label{fitting 1}
}  
In the upper left panel in Fig.~\ref{fig:virial}, we plot this function by blue, while the known virial coefficients are shown by red. 
Equation~(\ref{fitting 1}) implies that the radius of convergence of virial expansion~(\ref{dimensionless virial}) is $\eta_c^{}=1$, similar to the $D=1$ case.  
Besides, Eq.~(\ref{fitting 1}) also predicts an approximate analytic continuation of the pressure as    
\aln{
\beta p V_3^{}(a)\approx 1.28\times \mathrm{Li}_{-1.90}^{}(\eta)~,
\label{analytic continuation D=3}
} 
which has a branch cut in the unphysical density region $\eta\geq 1$.  
In the lower-left panel in Fig.~\ref{fig:virial}, we plot $\beta p V_3^{}(a)$ for various cases. 

By using the relation~(\ref{cluster by irreducible cluster}), we can also calculate the cluster integrals $\tilde{b}_l^{}\coloneq b_l^{}/V_3^{}(a)^{l-1}$ and find that it is well fitted by the following alternating series:
\aln{
\tilde{b}_l^{\mathrm{fit}}=(-1)^{l+1}0.0754\times (b_0^{})^{l}l^{-1.98}~,\quad \tilde{b}_0^{}=14.8.
}
They are also plotted in the upper-right panel in Fig.~\ref{fig:virial}. 
It is noteworthy that the $l$-dependence is similar to the large $D$ results~(\ref{large D pressure})(\ref{large D density}), suggesting that $D=3$ might be already regarded as large $D$.  
%Compared to the result~(\ref{bl by steepest descent}) by steepest descent method, the exponent of $l$ is found to be slightly different though it does not change the convergence radius:
%
In the present case, approximate analytic continuations of pressure and density as functions of $\tilde{z}\coloneq zV_3^{}(a)$ are given by
\aln{
\beta p V_3^{}(a)&\approx -0.0754\times \mathrm{Li}_{1.98}^{}(-\tilde{b}_0^{}\tilde{z})~,
\\
\eta &\approx -0.0754\times \mathrm{Li}_{0.98}^{}(-\tilde{b}_0^{}\tilde{z})~,
}
which should be compared to the large $D$ limit (\ref{large D pressure})(\ref{large D density}). 

Before concluding, an additional comment is necessary. 
In the above analysis, it was simply assumed that the fitted virial coefficients~(\ref{fitting 1}) are all positive for $k\geq 13$. 
However, as pointed out in the literatures~\cite{Clisby_2005,domb1989phase}, we cannot rule out a possibility that these coefficients may become negative.  
At present, a slight discrepancy between the known $\tilde{B}_{12}^{}$ and Eq.~(\ref{fitting 1}) might indicate a potential oscillation behavior  of the virial coefficients.
For illustration, we plot the ratio $\tilde{B}_k^{}/\tilde{B}_{k-1}^{}$ in the lower-right panel in Fig.~\ref{fig:virial}.  
More information on the higher-order virial coefficients is essential to improve the analysis. 

%______________________________________________________
\section{Conclusion}\label{sec:conclusion}

We have studied the asymptotic behavior of the virial series of the hard-sphere model. 
We have found that the known virial coefficients are well fitted by Eq.~(\ref{fitting 1}), leading to the approximate analytic continuation~(\ref{analytic continuation D=3}). 
These results imply that there is no indications of phase transition all the way up to the close-packing density $\eta_{\mathrm{max}}^{}\approx 0.74$ in the virial expansion approach. 
However, this result apparently contradicts the well established fact of the Alder phase transition. 
Additional information on the higher-order virial coefficients is  necessary to improve the analysis. 
%

%%%%%%%%%%%%%%%%%%%% ACKNOWLEDGMENTS %%%%%%%%%%%%%%%%%%%%
\section*{Acknowledgements} 
We would like to thank Takeru Yokota for the fruitful comments.    
%
%K.K. would like to thank Yukawa Institute for Theoretical Physics, Kyoto University for the support and the hospitality during his stay by the long term visiting program. 
This work is supported by KIAS Individual Grants, Grant No. 090901.

%%%%%%%%%%%%%%%%%%%%%%%%%%%Appendix%%%%%%%%%%%%%%%%%%%%%%%%%%%%%%%%%%%%%%%%
%\appendix 
%\section{Flow equations for higher-order vertices}

\bibliographystyle{TitleAndArxiv}
\bibliography{Bibliography}

\end{document}